\documentclass[12pt]{article}

\usepackage{graphicx}
\usepackage{color}
\usepackage{amsfonts}
\usepackage{amsmath}

\usepackage{amssymb}

\begin{document}
 
\newcommand{\al}{\mbox{$\alpha$}}
\newcommand{\be}{\mbox{$\beta$}}
\newcommand{\ep}{\mbox{$\epsilon$}}
\newcommand{\gam}{\mbox{$\gamma$}}
\newcommand{\sig}{\mbox{$\sigma$}}

\DeclareRobustCommand{\FIN}{%
  \ifmmode 
  \else \leavevmode\unskip\penalty9999 \hbox{}\nobreak\hfill
  \fi
  $\bullet$ \vspace{5mm}}

\newcommand{\calA}{\mbox{${\cal A}$}}
\newcommand{\calB}{\mbox{${\cal B}$}}
\newcommand{\calC}{\mbox{${\cal C}$}}

\newcommand{\muas}{\mbox{$\mu$-a.s.}}
\newcommand{\Nat}{\mbox{$\mathbb{N}$}}
\newcommand{\Rea}{\mbox{$\mathbb{R}$}}
\newcommand{\Prob}{\mbox{$\mathbb{P}$}}
\newcommand{\Read}{\mbox{$ \mathbb{R}^d$} }

\newcommand{\nin}{\mbox{$n \in \mathbb{N}$}}
\newcommand{\suc}{\mbox{$\{X_{n}\}$}}
\newcommand{\sucP}{\mbox{$\mathbb{P}_{n}\}$}}

\newcommand{\conv}{\rightarrow}
\newcommand{\convn}{\rightarrow_{n\rightarrow \infty}}
\newcommand{\convp}{\rightarrow_{\mbox{p}}}
\newcommand{\convs}{\rightarrow_{\mbox{a.s.}}}
\newcommand{\convw}{\rightarrow_w}
\newcommand{\convd}{\stackrel{\cal D}{\rightarrow}}

\newtheorem {Prop}{Proposition} [section]
 \newtheorem {Lemm}[Prop] {Lemma}
 \newtheorem {Theo}[Prop]{Theorem}
 \newtheorem {Coro}[Prop] {Corollary}
 \newtheorem {Nota}{Remark}[Prop]
 \newtheorem {Ejem}[Prop] {Example}
 \newtheorem {Defi}[Prop]{Definition}
 \newtheorem {Figu}[Prop]{Figure}
 \newtheorem {Table}[Prop]{Table}

\title{An { optimal transportation} approach for assessing almost stochastic order.\footnote{Research partially supported by the
Spanish Ministerio de Econom\'{\i}a y Competitividad y fondos FEDER, grants  
MTM2014-56235-C2-1-P and MTM2014-56235-C2-2.}}

\author{E. del Barrio$^{1}$, J.A. Cuesta-Albertos$^{2}$\\ and C. Matr\'an$^{1}$ \\
$^{1}$\textit{Departamento de Estad\'{\i}stica e Investigaci\'on Operativa and IMUVA,}\\
\textit{Universidad de Valladolid} \\ $^{2}$ \textit{Departamento de
Matem\'{a}ticas, Estad\'{\i}stica y Computaci\'{o}n,}\\
\textit{Universidad de Cantabria}}
\maketitle

\begin{abstract}
When stochastic dominance $F\leq_{st}G$ does not hold, we can improve agreement 
{ to stochastic order} by suitably { trimming} both { distributions}. 
In this work we consider { the $L_2-$Wasserstein distance, $\mathcal W_2$, to stochastic order of 
these trimmed versions}. Our characterization for that distance 
naturally leads to consider a $\mathcal W_2$-based index of disagreement with  
stochastic order, $\varepsilon_{\mathcal W_2}(F,G)$. We { provide asymptotic
results} allowing to test $H_0: \varepsilon_{\mathcal W_2}(F,G)\geq \varepsilon_0$ vs 
$H_a: \varepsilon_{\mathcal W_2}(F,G)<\varepsilon_0$, that, under rejection, 
would give statistical guarantee of almost stochastic dominance. 
{ We include a simulation study showing a good performance of the index under the normal model.}
\end{abstract}

\section{Introduction}

Let $P,Q$ be probability distributions on the real line 
with distribution functions (d.f.'s in the sequel) $F,G$, respectively.
Stochastic dominance of $Q$ over $P$, denoted $P\leq_{st}Q,$ is defined in terms 
of the d.f.'s by $F(x)\geq G(x)$ for every $x\in \Rea$   
(throughout we will also use the alternative notation $F\leq_{st}G$). 
The meaning of this relation is that { random outcomes produced by the second law tend to be
larger than those produce by the first one. We gain a 
better understanding of this stochastic order by considering a quantile representation.} For a d.f. $F$, 
the quantile function associated to $F$, that we will denote by $F^{-1}$, is defined by
$$F^{-1}(t)= \inf \{x: t\leq F(x)\}, \ t \in (0,1).$$
The following well-known statements (see e.g. \cite{Lehmann}) are equivalent to $F\leq_{st}G$:  
\begin{itemize}
\item[a)] There exist random variables $X,Y$ defined on some probability space $(\Omega,\sigma,\mu)$, 
with respective laws $P$ and $Q$ ($\mathcal L(X)=P, \mathcal L(Y)=Q$), { satisfying} $\mu(X\leq Y)=1$.
\item[b)] $F^{-1}(t)\leq G^{-1}(t)$ for every $t \in (0,1)$. 
\end{itemize}
Quantile functions (also called `monotone rearrangements' in other contexts) 
are characterized by $F^{-1}(t)\leq x$ if and only if $t\leq F(x)$. Therefore it 
is straightforward that, when considered as random variables defined on the unit 
interval with the Lebesgue measure $((0,1),\beta_{(0,1)},\ell)$, they satisfy 
$\mathcal L(F^{-1})=P, \mathcal L(G^{-1})=Q$. This representation shows that a) 
and b) are equivalent and, more importantly in the present setting, allows 
us to relate characteristics and measure agreement or disagreements with the stochastic 
order.

From the previous considerations it becomes clear that guaranteeing stochastic dominance,
$F \leq_{st} G$, should be the goal when comparing treatments or production schemes. 
However, { the rejection of $F \nleq_{st}G,$  
on the basis of { two} data samples is an ill posed statistical problem: 
As showed in \cite{Berger} and noted in \cite{Levy}, \cite{Davidson2007}, 
or \cite{Alv2015}, the `non-data test', namely the test which rejects with probability $\alpha$, regardless the data, 
is uniformly most powerful for  testing the nonparametric hypotheses $H_0: F \nleq_{st}G$ vs $H_a: F \leq_{st}G$.}
This fact motivates  recent research looking for suitable indices measuring `almost' or `approximate' versions of  
stochastic dominance. Here, suitability of an index must be understood  in terms of computability and interpretability, 
{ but also in terms of statistical performance}.
Usually, as already suggested in 
a general context in \cite{Hodges}, such measures of nearness involve the use of some kind of distance to the null. 
This will also be the approach here, with the choice of the $L_2$-Wasserstein distance between probabilities. 
For  $P,Q$ in the set $\mathcal{F}_2(\Read)$ of Borel probabilities on $\Read$ with finite second order moments, 
this distance is defined as
$$\mathcal W_2(P,Q):= \min\left\{\int\|x-y\|^2d\nu(x,y), \nu \in\mathcal{F}_2(\Read\times\Read) \mbox{ with marginals } 
P,Q\right\}^{1/2}.$$
In the univariate case, $\mathcal W_2$ equals the $L_2$-distance between quantile functions, namely,
\begin{equation}\label{distancia}
\mathcal W_2(P,Q)=\left(\int_0^1 |F^{-1}(t)-G^{-1}(t)|^2dt \right)^{1/2}.
\end{equation}
Statistical applications based on optimal transportation, and particularly on the $L_2$ version, are receiving considerable 
attention in recent times (see e.g. \cite{Bois15}, \cite{Carlier}, \cite{Depth}, \cite{Rippl} or \cite{Alv2017}). 
We should mention here our papers \cite{Alv2011} and \cite{Alv2011b}, dealing with similarity of 
distributions (as a relaxation of homogeneity) through this distance, 
and also \cite{Alv2015} (and \cite{Alv2014}) which introduced 
an index of disagreement from stochastic dominance based on the idea of similarity. 
The key to this index  is the existence, 
for a given (small enough)  $\pi,$ of mixture decompositions 
\begin{equation}\label{modelocontaminadotwosample1}
\left\{
\begin{matrix}
F & = & (1-\pi)\tilde F+\pi H_F\\
& &\\
G & = & (1-\pi)\tilde G+\pi H_G,
\end{matrix}
\right.
\quad \mbox{ for some d.f.'s } \tilde F, H_F, \tilde G, H_G \mbox{ such that } \tilde F\leq_{st} \tilde G.
\end{equation}
{ If model (\ref{modelocontaminadotwosample1}) holds then it means that stochastic order holds after removing 
contaminating $\pi$-fractions from each population.
The minimum $\pi$ compatible with (\ref{modelocontaminadotwosample1}), 
denoted by $\pi(F,G)$, can then be taken as a measure of deviation from stochastic order, 
see \cite{Alv2015} for details.} 
We would like to emphasize here that the analysis in \cite{Alv2015} is based on the connection between
contamination models and  trimmed probabilities. We recall that an $\alpha$-trimming of a 
probability, $P$, is any other probability, say $\tilde P$, such that 
$$\tilde P(A)=\int_A \tau dP \quad \mbox{ for everys event } A$$
for some function $\tau$ taking values in $[0,\frac 1 {1-\alpha}]$. Like the trimming methods,
commonly used in Robust Statistics, consisting of removing disturbing observations, 
the function $\tau$  allows to discard or downplay the influence of some regions on the sample space. 
On the real line, writing $\mathcal{R}_\alpha(F)$ for the set of trimmings of $F$, it turns out (see \cite{Alv2015}) that 
\begin{equation}\label{trimmingcontam}
F =(1-\alpha)\tilde F+\alpha H_F \mbox{ for some d.f.'s } \tilde F, H_F
\mbox{ if and only if } \tilde F\in \mathcal{R}_\alpha(F).
\end{equation}

{ The contaminated stochastic order model (\ref{modelocontaminadotwosample1}) can also be recast
in terms of trimmings. If we denote 
$$\mathcal{F}_{st}:=\{(H_1,H_2)\in \mathcal{F}_2\times \mathcal{F}_2:\quad H_1\leq_{st} H_2\},$$
then, for $F,G\in\mathcal{F}_2$, (\ref{modelocontaminadotwosample1}) holds if and only if
\begin{equation}\label{equivstpi}
(\mathcal{R}_\pi(F)\times \mathcal{R}_\pi(G)) \cap \mathcal{F}_{st} \ne \emptyset
\end{equation}
or, equivalently (this follows from compactness of $\mathcal{R}_\pi(F)\times \mathcal{R}_\pi(G)$ with
respect to $d_2$; we omit details), if and only if
\begin{equation}\label{equivstpi2}
d_2(\mathcal{R}_\pi(F)\times \mathcal{R}_\pi(G), \mathcal{F}_{st})=0,
\end{equation}
where $d_2$ denotes the metric on the set $\mathcal{F}_2\times \mathcal{F}_2$ given by
$$d_2((F_1,F_2),(G_1,G_2))=\sqrt{\mathcal{W}_2^2(F_1,G_1)+\mathcal{W}_2^2(F_2,G_2)}$$
and, for $A,B\subset\mathcal{F}_2\times \mathcal{F}_2$, $d_2(A,B)=\inf_{a\in A,b\in B} d_2(a,b)$.

For fixed $\pi$, $d_2(\mathcal{R}_\pi(F)\times \mathcal{R}_\pi(G), \mathcal{F}_{st})$ 
can be used as a measure of deviation from the contaminated stochastic order model
(\ref{modelocontaminadotwosample1}). In this work we obtain a simple explicit 
characterization of this measure (see Theorem \ref{minimadistancia} below) 
that could be used  for statistical purposes. Later, we use this characterization 
to introduce a new index, $\varepsilon_{\mathcal{W}_2}$, see (\ref{theindex}), to evaluate disagreement 
with respect to the (non-contaminated) stochastic order. 
We also provide asymptotic theory (Theorem \ref{asymptotics}) about the behavior of this index, that allows addressing the goal of statistical assessment 
of $\varepsilon_{\mathcal{W}_2}$-almost stochastic dominance. This index has some similarity 
with that proposed in \cite{Levy} for which, 
in contrast, asymptotics are not available. 

The remaining sections of this work are organized as follows. Section 2 presents the announced results, introduces
the new index $\varepsilon_{\mathcal{W}_2}$ and discusses its application in the statistical assessment of almost stochastic order.
This includes an illustration of the meaning of the index in the case of normal distributions and a small simulation study. Finally, 
the more technical proof of Theorem \ref{asymptotics} is given in an Appendix.}


\section{Main results}\label{results}

 A fortunate fact that eases the use of trimming in the stochastic dominance setting is that the set $\mathcal{R}_\alpha(F)$ has a minimum and a maximum for the stochastic order. Moreover both can be easily characterized as follows (see Proposition 2.3 in \cite{Alv2015}).
 \begin{Prop}\label{prop2}
Consider a distribution function $F$ and $\pi\in [0,1)$. Define the d.f.'s
\begin{eqnarray*}
F^\pi(x) = \max({\textstyle \frac 1 {1-\pi}} (F(x) - \pi),0)\quad \mbox{and}\quad
F_\pi(x) =
\min({\textstyle \frac 1 {1-\pi}} F(x),1). 
\end{eqnarray*}
Then $F^\pi,F_\pi\in \mathcal{R}_\pi (F)$ and any other $\tilde{F}\in\mathcal{R}_\pi (F)$
satisfies
$F_\pi\leq_{st}\tilde{F}\leq_{st} F^\pi$.
\end{Prop}

Recalling the characterization of the stochastic order in terms of  quantile functions, 
a simple computation shows that the associated quantile functions are
\begin{equation}\label{notacioncuantiles}
(F_{\pi})^{-1}(t)=F^{-1}((1-\pi)t),\quad (F^{\pi})^{-1}(t)=F^{-1}(\pi+(1-\pi)t),\quad 0<t<1,
\end{equation}
so we can restate this proposition in the following new way.

\begin{Prop}\label{minmaxquantile}
If $\tilde F \in \mathcal{R}_\pi(F)$, then its quantile function
satisfies
\begin{equation}\label{quantilenvelope}
F^{-1}((1-\pi)t)\leq \tilde F^{-1}(t)\leq F^{-1}(\pi+(1-\pi)t), \quad 0<t<1.
\end{equation}
\end{Prop}

We can use equation (\ref{quantilenvelope}) for proving our next result, the announced
characterization for $d_2(\mathcal{R}_\pi(F)\times \mathcal{R}_\pi(G), \mathcal{F}_{st})$, a quantity that measures
deviation from the contaminated stochastic order model (\ref{modelocontaminadotwosample1}).
We keep the notation in (\ref{notacioncuantiles}) and
define 
$$(L_\pi)^{-1}(t)=\left\{
\begin{matrix}
(F_\pi)^{-1}(t)  & \mbox{ if } (F_\pi)^{-1}(t)\leq (G^\pi)^{-1}(t) \\
\frac 1 2 ((F_\pi)^{-1}(t)+(G^\pi)^{-1}(t)) & \mbox{ if } (F_\pi)^{-1}(t)>(G^\pi)^{-1}(t)
\end{matrix}
\right.
$$
$$
(U_\pi)^{-1}(t)=\left\{
\begin{matrix}
(G^\pi)^{-1}(t)  & \mbox{ if } (F_\pi)^{-1}(t)\leq (G^\pi)^{-1}(t) \\
\frac 1 2 ((F_\pi)^{-1}(t)+(G^\pi)^{-1}(t)) & \mbox{ if } (F_\pi)^{-1}(t)>(G^\pi)^{-1}(t).
\end{matrix}
\right.
$$
\begin{Theo}\label{minimadistancia}
With the above notation, if $F$ and $G$ are distribution functions with finite second moment the 
$L_\pi^{-1}$, $U_{\pi}^{-1}$ are the quantile functions of a pair $(L_\pi,U_\pi)\in\mathcal{F}_{st}$.
Furthermore, if we denote $x_+=\max(x,0)$, 
\begin{eqnarray*}
d_2(\mathcal{R}_\pi(F)\times \mathcal{R}_\pi(G), \mathcal{F}_{st})&=&d_2((F_\pi,G^{\pi}),(L_\pi, U_\pi))\\
&=&\sqrt{\frac 1 2 \int_0^1 (F^{-1}((1-\pi)t)-G^{-1}(\pi+(1-\pi)t))_+^2 dt}.
\end{eqnarray*}
\end{Theo}

\medskip
\noindent\textbf{Proof.} To see that $(L_\pi)^{-1}$ is a quantile function we note that
$$(L_\pi)^{-1}(t)=\min((F_\pi)^{-1}(t), {\textstyle \frac 1 2} ((F_\pi)^{-1}(t)+(G^\pi)^{-1}(t))).$$
This shows that $(L_\pi)^{-1}$ is nondecreasing and left continuous, hence a quantile function.
That $L_\pi$ has finite second moment follows from the elementary
bounds
$$-(|(F_\pi)^{-1}(t)|+|(G^\pi)^{-1}(t)|) \leq (L_\pi)^{-1}(t)\leq (F_\pi)^{-1}(t).$$
A similar argument works for $U_\pi$. Obviously $L_\pi \leq_{st} U_\pi$ and, therefore,
$(L_\pi,U_\pi)\in\mathcal{F}_{st}$. Now, for any $(U_1,U_2)\in \mathcal{R}_\pi(F)\times \mathcal{R}_\pi(G)$
and $(V_1,V_2)\in \mathcal{F}_{st}$ we have $U_1^{-1}(t)\geq (F_\pi)^{-1}(t)$, $U_2^{-1}(t)\leq (G^\pi)^{-1}(t)$,
$V_1^{-1}(t)\leq V_2^{-1}(t)$.
We define $A_{\pi}:=\{t\in(0,1):\, (F_\pi)^{-1}(t)>(G^\pi)^{-1}(t)\}$. Then
\begin{eqnarray*}
d_2((U_1,U_2),(V_1,V_2))&=&\int_0^1 ( (U_1^{-1}(t)-V_1^{-1}(t))^2+ (U_2^{-1}(t)-V_2^{-1}(t))^2 )dt\\
&\geq & \int_{A _{\pi}}((U_1^{-1}(t)-V_1^{-1}(t))^2+ (U_2^{-1}(t)-V_2^{-1}(t))^2 )dt\\
&\geq & \int_{A _{\pi}} (((F_\pi)^{-1}(t)-(L_\pi)^{-1}(t))^2+((G^\pi)^{-1}(t)-(U_\pi)^{-1}(t))^2) dt\\
&=& \int_0^1 (((F_\pi)^{-1}(t)-(L_\pi)^{-1}(t))^2+((G^\pi)^{-1}(t)-(U_\pi)^{-1}(t))^2) dt \\
&=& d_2((F_\pi,G^{\pi}),(L_\pi, U_\pi)),
\end{eqnarray*}
where the last lower bound is just the trivial fact that if $f>g$, then the minimum value $\min_{a,b,c,d}(a-b)^2+(c-d)^2$, for $a\geq f, c\leq g, b\leq d$ is just attained at  $a=f$, $c=g$, $b=d=\frac{f+g}2$. To complete
the proof we note that 
\begin{eqnarray*}
d_2((F_\pi,G^{\pi}),(L_\pi, U_\pi))&=&\frac 1 2 \int_{A _{\pi}} ((F_\pi)^{-1}(t)-(G^\pi)^{-1}(t))^2 dt\\
&=& \frac 1 2 \int_0^1 (F^{-1}((1-\pi)t)-G^{-1}(\pi+(1-\pi)t))_+^2 dt
\end{eqnarray*}\FIN

Particularizing for $\pi=0$, Theorem \ref{minimadistancia} shows that the distance $d_2$ between the 
pair $(F,G)$ and the set $\mathcal F_{st}$ is attained  at the pair $(L_0,U_0)\in \mathcal F_{st}$ associated to 
the quantile functions $L_0^{-1}=\inf \{F^{-1},(F^{-1}+G^{-1})/2\}$ and $U_0^{-1}=\sup \{G^{-1},(F^{-1}+G^{-1})/2\}$. 
Moreover, $d_2^2((F,G),\mathcal{F}_{st})=\frac 1 2 \int_0^1 (F^{-1}(t)-G^{-1}(t))_+^2 dt$. 
Avoiding the factor $1/2$, this is just the part of $\mathcal W_2^2(F,G)$ due to the violation of stochastic dominance. 
Therefore, for distinct d.f.'s $F,G$, according to the notation $A_0=\{t\in (0,1): F^{-1}(t)>G^{-1}(t)\}$, the quotient 
\begin{equation}\label{theindex}
\varepsilon_{\mathcal W_2}(F,G):=\frac{\int_{A_0} (F^{-1}(t)-G^{-1}(t))^2 dt}{ \mathcal W_2^2(F,G)}
\end{equation}
can be considered as a normalized index of such violation. 
It satisfies $0\leq \varepsilon_{\mathcal W_2}(F,G)\leq 1,$ with the extreme values 0 and 1 
corresponding, respectively, to perfect stochastic dominance of $G$ over $F$ and vice-versa. 
We notice that \cite{Levy}, following a very different motivation, introduced a related index 
consisting in the quotient ${\int_{F<G} (G(x)-F(x)) dx}/{ \|F-G\|_1},$ where $\|-\|_1$ is the $L_1-$ norm with respect to the Lebesgue measure on the line.
 
The index $\varepsilon_{\mathcal W_2}(F,G)$ can be estimated by its sample counterpart $\varepsilon_{\mathcal W_2}(F_n,G_m)$, when $F_n$ and $G_m$ are the sample d.f.'s associated to  independent samples respectively obtained from $F$ and $G$. The following theorem gives the mathematical background for such task.
 
\begin{Theo}\label{asymptotics}
Let $F, G$ be distinct d.f.'s in $\mathcal F_2$ and assume $n,m\to \infty$ with 
$\frac n{n+m}\to \lambda \in (0,1)$.  If $F_n$ and $G_m$ are the sample d.f.'s based on 
independent samples of $F$ and $G$, then  $\varepsilon_{\mathcal W_2}(F_n,G_m)\to 
\varepsilon_{\mathcal W_2}(F,G)$ a.s.. If, additionally, $F$ and $G$ have bounded 
convex supports, then 
\begin{equation}\label{dosmuestras}
\sqrt{\frac{mn}{m+n}}\left(\varepsilon_{\mathcal W_2}(F_n,G_m)-
\varepsilon_{\mathcal W_2}(F,G)\right) \convw N(0,\sigma^2_\lambda(F,G)),
\end{equation}
where 
$$\sigma^2_\lambda(F,G)=\frac 1 {\mathcal{W}_2^8(F,G)}[(1-\lambda)\mbox{\em Var}(u_-(X)) +\lambda \mbox{\em Var} (u_+(Y))],$$
$u_+(x)=\int_0^x 2(s-G^{-1}(F(s)))_+ds$, $u_-(x)=\int_0^x 2(s-G^{-1}(F(s)))_-ds$ and 
$X$ and $Y$ are independent r.v.'s with d.f.'s $F$ and $G$, respectively. 
\end{Theo}

{ A critical analysis of the problem of assessing improvement in a treatment comparison setup
from the perspective of stochastic dominance is given in \cite{Alv2017b}.
It is argued  there that under, say, normality assumptions, improvement with the new treatment 
is often assessed through a one sided test for the mean, while the really interesting test would be 
that of $F\nleq_{st}G$ vs $F\leq_{st}G$. Since, as argued in the Introduction, this is 
not a feasible statistical task, we emphasized there 
on the alternative, feasible goal of testing that slightly relaxed versions of stochastic 
dominance hold. In the present setting, such a strategy leads to consider the problem of testing, at a given 
confidence level, $H_0: \varepsilon_{\mathcal W_2}(F,G)\geq \varepsilon_0$ vs $H_a: 
\varepsilon_{\mathcal W_2}(F,G)<\varepsilon_0$, where $\varepsilon_0$ is a small enough prefixed 
amount of disagreement with the stochastic order.}

Following the scheme in \cite{Alv2015} and \cite{Alv2017b}, from  the asymptotic normality obtained 
in Theorem \ref{asymptotics} we propose to reject $H_0$ if
\begin{equation}\label{proposal}
\textstyle{\sqrt{\frac{nm}{n+m}}
(\varepsilon_{\mathcal W_2}(F_n,G_m)-\varepsilon_0) < \hat{\sigma}_{n,m}\Phi^{-1}(\alpha)},
\end{equation}
where $\hat{\sigma}_{n,m}$ is an estimator of $\sigma_\lambda(F,G)$ (for example a bootstrap estimator). 
This rejection rule provides a consistent test of asymptotic level $\alpha$. 
Also,  
\begin{equation}\label{upperconfidencedirectepsilon}
\hat{U}:=\varepsilon_{\mathcal W_2}(F_n,G_m)-\textstyle \sqrt{\frac{n+m}{nm}} \hat{\sigma}_{n,m}\Phi^{-1}(\alpha)
\end{equation}
provides an upper confidence bound for  $\varepsilon_{\mathcal W_2}(F,G)$ with asymptotic level $1-\alpha$.

\begin{figure}[ht] 
\begin{center}
\includegraphics[scale=.4]{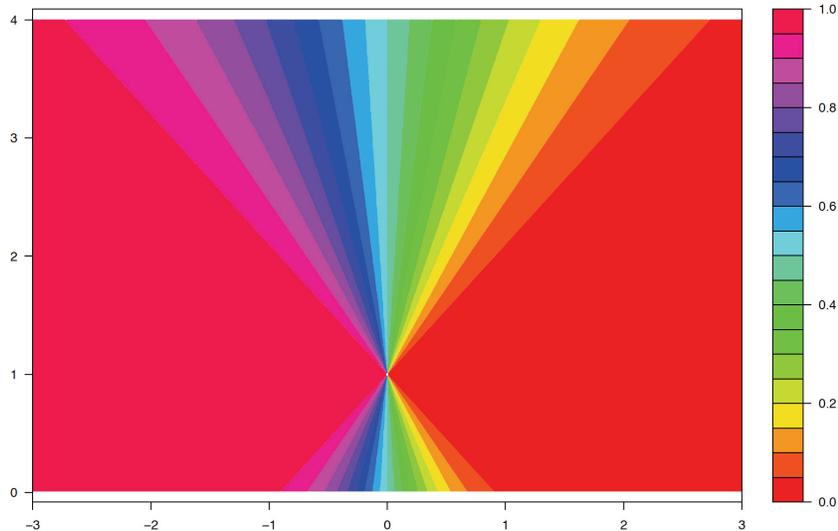} 
\vspace{-10mm}
\caption{Contour-plot of $\varepsilon_{\mathcal W_2}(N(0,1),N(\mu,\sigma^2))$ as in (\ref{theindex}) for 
different values of $\mu$ (X-axis) and $\sigma$ (Y-axis)} 
 \label{figura}
\end{center}
\end{figure}

{ Let us take now a closer look at the $\varepsilon_{\mathcal W_2}$ index for distributions in 
a location-scale family. For simplicity, we focus on normal laws. It is an elementary fact that $\varepsilon_{\mathcal W_2}$
is invariant to changes in location and scale and we can, consequently, resctrict ourselves to the analysis of 
Therefore we can obtain the values of  $\varepsilon_{\mathcal W_2}(N(0,1),N(\mu,\sigma^2))$, $\mu\in\mathbb{R}, \sigma>0$. 
Moreover, it is easy to see that $\varepsilon_{\mathcal W_2}$ is constant when $(\mu,\sigma)$ moves 
along directed rays from $(0,1)$. This fact is showed in figure \ref{figura}. We see that $\mu>0$ corresponds to 
$\varepsilon_{\mathcal W_2}(N(0,1),N(\mu,\sigma^2))<\frac 1 2$, with $\varepsilon_{\mathcal W_2}(N(0,1),N(\mu,1))=0$,
but the index can be made arbitrarily close to $\frac 1 2$ by taking $\sigma$ large enough.} 

Finally, we present in Table \ref{Tabla} some simulations showing 
the performance of the proposed nonparametric procedure. We see the observed 
rejection rates for the test (\ref{proposal}). In our simulations we have taken $F=N(0,1)$ and
$G=N(\mu,\sigma^2)$ for several choices of $\mu,\sigma$. We show also the rejection rates
based on a natural competitor, the parametric maximum likelihood estimator 
$\hat\varepsilon_{\mathcal W_2}:=\varepsilon_{\mathcal W_2}(F_{N(\bar X_n,S_X^2)},F_{N(\bar Y_m,S_Y^2)})$. {
This estimator is, of course, highly nonrobust and useless in practice without the a priori knowledge that $F$ and $G$ are
normal, but we use it here as a benchmark. We see a reasonable amount of agreement of the rejection frequencies to the
nominal level of the test, even if it is slightly liberal for $\sigma$ close to one and small $\varepsilon_0$, but the nonparamentric
procedure does not perform worse than the parametric benchmark. We also see that it is possible to get statistical evidence
that almost stochastic order does hold. For instance, for $\mu=.697$, $\sigma=1.5$ (true $\varepsilon_{\mathcal W_2}=0.01$)
sizes $n=m=1000$ suffice to conclude that $\varepsilon_{\mathcal W_2}<0.05$ with probability close to $0.93$.}

\begin{table}[ht]\label{Tabla}
\caption{\label{Tabla.2} Rejection rates for 
$\varepsilon_{\mathcal W_2}(N(0,1),N(\mu,\sigma^2)) \geq \varepsilon_0$ 
at level $\alpha =.05$ along 1,000 simulations. Upper 
(resp. lower) rows show results for nonparametric (resp. parametric) 
comparisons. For each $\sigma$, $\mu$ is chosen to make
$\varepsilon_{\mathcal W_2}(N(0,1),N(\mu,\sigma^2))=$ 0.01, 0.05 and 0.10 (first, second and third columns, resp.).}
\begin{center}
\begin{tabular}{|cr|ccc|ccc|ccc|}
\hline
& &\multicolumn{3}{|c}{$\sigma =1.1$}&\multicolumn{3}{|c|}{$\sigma = 1.5$}&\multicolumn{3}{|c|}{$\sigma = 2$}
\\
&Sample&\multicolumn{3}{|c}{$\mu$}&\multicolumn{3}{|c}{$\mu$}&\multicolumn{3}{|c|}{$\mu$}
\\
$\varepsilon_0$&\multicolumn{1}{c|}{size}&$.139$ &$.091$ &$.068$ &$.697$ &$.455$ &$.341$ &$1.395$ &$ .909$ &$.683$ 
\\
\hline
.01 & 100 & .000 & .000 & .000 & .053 & .007 & .000 & .180 & .009 & .004
\\
 &  & .000 & .000 & .000 & .062 & .006 & .000 & .112 & .003 & .000
\\
\cline{2-11}
 & 1000 & .004 & .000 & .000 & .086 & .000 & .000 & .116 & .000 & .000
\\
 &  & .036 & .002 & .000 & .086 & .000 & .000 & .086 & .000 & .000
\\
\cline{2-11}
 & 5000 & .014 & .000 & .000 & .084 & .000 & .000 & .077 & .000 & .000
\\
 &  & .078 & .003 & .000 & .086 & .000 & .000 & .060 & .000 & .000
\\
\hline
.05 & 100 & .013 & .004 & .004 & .321 &  .060 & .019 & .677 &  .138 & .028
\\
 &  & .017 & .007 & .004 & .382 &  .064 & .027 & .690 &  .086 & .017
\\
\cline{2-11}
 & 1000 & .101 & .017 & .004 & .929 &  .088 & .003 & .999 &  .101 & .000
\\
 &  & .219 & .041 & .015 & .982 &  .087 & .002 & 1.000 &  .085 & .000
\\
\cline{2-11}
 & 5000 & .488 & .056 & .009 & 1.000 &  .067 & .000 & 1.000 &  .070 & .000
\\
 &  & .704 & .099 & .009 & 1.000 &  .069 & .000 & 1.000 &  .057 & .000
\\
\hline

.10 & 100 & .034 &  .017 & .006 & .608 &   .210 & .092 & .930 &  .402 & .148
\\
 &  & .040 &  .022 & .009 & .658 &   .205 & .073 & .941 &  .364 & .109
\\
\cline{2-11}
 & 1000 & .267 &  .082 & .020 & 1.000 &   .545 & .076 & 1.000 &  .861 & .096
\\
 &  & .431 &  .132 & .047 & 1.000 &   .642 & .076 & 1.000 &  .928 & .084
\\
\cline{2-11}
 & 5000 & .867 &  .246 & .058 & 1.000 &   .970 & .056 & 1.000 & 1.000 & .078
\\
 &  & .960 &  .356 & .087 & 1.000 &   .994 & .058 & 1.000 & 1.000 & .069
\\
\hline
\end{tabular}
\end{center}
   \end{table}

\section{Appendix}

We prove here central limit theorems for the index $\varepsilon_{\mathcal{W}_2}$ in (\ref{theindex}). 
We will assume that $U_1,\ldots,U_n, V_1,\ldots,V_m$ are i.i.d. random variables, 
uniformly distributed on $(0,1)$. We consider independent samples i.i.d. $X_1,\ldots,X_n$ and
$Y_1,\ldots,Y_m$ such that the d.f. of the $X_i$ and the $Y_j$ are $F$ and $G$, respectively.
We note that, without loss of generality, we can assume that the $X_i$ and $Y_j$ are generated 
from the $U_i$ and the $V_j$ through  $X_i=F^{-1}(U_i)$, $Y_j=G^{-1}(V_j)$.
We write $F_n$, $G_m$, $H_{n,1}$ and $H_{m,2}$ for the empirical d.f.'s on the $X_i$, the $Y_j$, $U_i$ and the $V_j$, respectively.
Note that, in particular, $F_n^{-1}(t)=F^{-1}(H_{n,1}^{-1}(t))$, $G_m^{-1}(t)=G^{-1}(H_{m,2})$. Finally, $\alpha_{n,1}$ and
$\alpha_{m,2}$ will denote the empirical processes associated to the $U_i$ and the $Y_j$, namely, $\alpha_{n,1}(t)=\sqrt{n}(H_{n,1}(t)-t)$, $0\leq t\leq 1$,
and similarly for $\alpha_{m,2}$ and we will write $\alpha_{n,1}(h)$ instead of $\int_0^1 h(t)d\alpha_{n,1}(t)$.

We introduce the statistics $S_n=\int_0^1 (F_n^{-1}-G^{-1})^2$, $S_n^+=\int_0^1 (F_n^{-1}-G^{-1})_+^2$, 
$S_n^-=\int_0^1 (F_n^{-1}-G^{-1})_-^2$, and write $S$, $S^+$, $S_-$ for the corresponding population counterparts. 
Note that, to ensure that $S$ is finite, $F$ and $G$ should have, at least,  finite
second moments. However, to simplify the arguments our proof will require bounded supports. We set 
$$T_n=\sqrt{n}(S_n-S),\quad T_n^+=\sqrt{n}(S^+_n-S^+),\quad T_n^-=\sqrt{n}(S_n^--S^-),$$
$c(x)=2x$, $c_+(x)=2x_+$, $c_-(x)=2x_-$ and define
\begin{equation}\label{funcionesv}\textstyle
v(t)=\int_0^{F^{-1}(t)}c(s-G^{-1}(F(s)))ds
\end{equation}
and similarly $v_+$ and $v_-$ replacing $c$ with $c_+$ and $c_-$, respectively. Observe that $v=v_+-v_-$.
We this notation we have the following result.
 
\begin{Theo}\label{casoL2}
If $F$ and $G$ have bounded support and $G^{-1}$ is continuous on $(0,1)$ then 
$$T_n= \alpha_{n,1}(v)+o_P(1),\quad T^+_n=\alpha_{n,1}(v_+)+o_P(1),\quad  T^-_n=\alpha_{n,1}(v_-)+o_P(1).$$
\end{Theo}

\medskip
\noindent
\textbf{Proof.} We assume that $|F^{-1}(t)|\leq M $, $|G^{-1}(t)|\leq M$ for 
all $t\in (0,1)$ and some $M>0$. 
The continuity and boundedness assumption on $G^{-1}$ allows us to assume that $G^{-1}$ is a continuous
function on $[0,1]$, hence, uniformly coutinuous and its modulus of continuity, 
$$\textstyle\omega(\delta)=\sup_{|t_1-t_2|\leq \delta} |G^{-1}(t_1)-G^{-1}(t_2)|,$$
satisfies $\omega(\delta)\to 0$ as $\delta\to 0$. 
It is convenient at this point to note that $T_n$ is a function of the $U_i$ and also of $F$ and we stress this fact
writing $T_n(F)$ instead of $T_n$ in this proof, and the same for $T_n^+$ and $T_n^-$. Similarly, we set 
$\tilde{T}_n(F)=\alpha_{n,1}(v)$, $\tilde{T}_n^+(F)=\alpha_{n,1}(v_+)$, $\tilde{T}_n^-(F)=\alpha_{n,1}(v_-)$.
We claim now that
\begin{equation}\label{linearization}\textstyle
E|T_n(F)-\tilde{T}_n(F)|^2\leq 16 M^2 E\Big(\|\alpha_{n,1}\|^2 \omega^2\Big(\frac{\|\alpha_{n,1}\|}{\sqrt{n}} \Big)\Big),
\end{equation}
where $\|\alpha_{n,1}\|=\sup_{0\leq t\leq 1} |\alpha_{n,1}(t)|$. To check this, let us assume first that $F$ is finitely supported,
say on $-M\leq x_1<\ldots<x_k\leq M$ with $F(x_j)=s_j$, $j=1,\ldots,k$ This means that $F^{-1}(t)=x_i$ if $s_{i-1}<t\leq s_i$ (we set $s_0=0$ for convenience)
and we have
\begin{eqnarray*}\textstyle
\int_0^1(F^{-1}-G^{-1})^2&=&\textstyle\sum_{i=1}^k \int_{s_{i-1}}^{s_i}(x_i-G^{-1}(t))^2dt= \int_{0}^{1}(x_k-G^{-1}(t))^2dt\\
&&\textstyle-\sum_{i=1}^{k-1} \int_{0}^{s_i} \big[(x_{i+1}-G^{-1}(t))^2-(x_i-G^{-1}(t))^2\big]dt\\
&=&\textstyle \int_{0}^{1}(x_k-G^{-1}(t))^2dt -\sum_{i=1}^{k-1} \int_{0}^{s_i} \big[\int_{x_i}^{x_{i+1}}c(s-G^{-1}(t))ds\big]dt.
\end{eqnarray*}
A similar expression holds for $\int_0^1(F_n^{-1}-G^{-1})^2$ replacing $s_i$ with $H_{n,1}(s_i)$ and we see that
$$\textstyle T_n(F)=-\sqrt{n} \sum_{i=1}^{k-1}\int_{s_i}^{H_{n,1}(s_i)}\Big(\int_{x_i}^{x_{i+1}} c(s-G^{-1}(t))ds\Big) dt.$$
We can argue analogously to check that
\begin{eqnarray*}
\tilde{T}_n(F)&=&\textstyle-\sum_{i=1}^{k-1}\alpha_{n,1}(s_i)\Big(\int_{x_i}^{x_{i+1}} c(s-G^{-1}(s_i))ds\Big) \\
&=&\textstyle-\sqrt{n} \sum_{i=1}^{k-1}\int_{s_i}^{H_{n,1}(s_i)}\Big(\int_{x_i}^{x_{i+1}} c(s-G^{-1}(s_i))ds\Big) dt.
\end{eqnarray*}
Hence, we see that
\begin{eqnarray*}
|T_n(F)-\tilde{T}_n(F)|&\leq &  \textstyle 2  \sum_{i=1}^{k-1} |\alpha_{n,1}(s_i)| (x_{i+1}-x_i) \omega\Big(\frac{\|\alpha_{n,1}\|}{\sqrt{n}} \Big)\\
&\leq & \textstyle 2 \|\alpha_n\| (x_{k}-x_1) \omega\Big(\frac{\|\alpha_{n,1}\|}{\sqrt{n}} \Big)\leq 4M \|\alpha_n\| \omega\Big(\frac{\|\alpha_{n,1}\|}{\sqrt{n}} \Big)
\end{eqnarray*}
and (\ref{linearization}) follows. For general $F$ take finitely supported $F_m$ such that $\hat F_m\to_w F$, $\hat F_m$ supported in $[-M,M]$. 
Then, for fixed $n$, $E|T_n(\hat F_m)-T_n(F)|^2\to 0$ and $E|\tilde{T}_n(\hat F_m)-\tilde{T}_n(F)|^2\to 0$ as $m\to\infty$. As a consequence, we conclude that 
(\ref{linearization}) holds also in this case.

Now, by the Dvoretzky-Kiefer-Wolfowitz inequality (see \cite{Massart1990}) we have
$P(\|\alpha_{n,1}\|>t)\leq 2 e^{-2t^2}$, $t>0$. This entails that $\|\alpha_{n,1}\|^2$ is uniformly
integrable and also that $\omega\Big(\frac{\|\alpha_{n,1}\|}{\sqrt{n}} \Big)$ vanishes in probability. Since, on the other hand,
$\omega^2\Big(\frac{\|\alpha_{n,1}\|}{\sqrt{n}} \Big)\|\alpha_{n,1}\|^2\leq M^2 \|\alpha_{n,1}\|^2$ we conclude that
\begin{equation}\label{limite0} \textstyle
E\Big(\|\alpha_{n,1}\|^2 \omega^2\Big(\frac{\|\alpha_{n,1}\|}{\sqrt{n}} \Big)\Big)\to 0
\end{equation}
as $n \to \infty$ and this proves the first claim in the Theorem. For the others, we can argue as above to see
that (\ref{linearization}) also holds if we replace $T_n(F)$ and $\tilde{T}_n(F)$ with the corresponding pairs $T_n^+(F)$ and $\tilde{T}_n^+(F)$
or $T_n^-(F)$ and $\tilde{T}_n^-(F)$. This completes the proof.
\FIN

From Theorem \ref{casoL2} we obtain a CLT for the one-sample empirical version of $\varepsilon_{\mathcal{W}_2}$.
\begin{Coro}\label{indexL2}
If $F$ and $G$ have bounded support and $G^{-1}$ is continuous, then 
$$\sqrt{n} (\varepsilon_{\mathcal{W}_2}(F_n,G)-\varepsilon_{\mathcal{W}_2}(F,G))\to_w N(0,\sigma^2)$$
with  $\sigma^2= \frac {\mbox{\em Var}(v_-(U))} {\mathcal{W}_2^8(F,G)},$ 
$v_-$ as in (\ref{funcionesv}) and $U$ a uniform r.v. on $(0,1)$.
\end{Coro}

\medskip
\noindent
\textbf{Proof.}  Observe that $\sqrt{n}(\varepsilon_{\mathcal{W}_2}(F_n,G)-\varepsilon_{\mathcal{W}_2}(F,G))= \sqrt{n} (\frac{S_n^+}{S_n}- \frac{S^+}{S})=\frac{1}{S S_n}
(T_n^+-T_n)$. From Theorem \ref{casoL2}, $(T_n^+-T_n)=\alpha_{n,1}(v_+-v)+o_P(1)=-\alpha_{n,1}(v_-)+o_P(1)$, while $S_n\to S$ a.s.\FIN

\begin{Nota}\label{Notadosmuestras}{\em
For the two-sample analogue of Corollary \ref{indexL2} it is important to observe that the conclusion of Theorem \ref{casoL2}
remains true if we replace $T_n$ by $\hat{T}_{n,m}:=\sqrt{n}(\int_0^1 (F_n^{-1}-G_m^{-1})^2-\int_0^1 (F^{-1}-G_m^{-1})^2)$ and $m\to\infty$.
In fact, in the finitely supported case, keeping the notation in the proof of Theorem \ref{casoL2}, we have
$$\textstyle\hat{T}_{n,m}=-\sqrt{n} \sum_{i=1}^{k-1}\int_{s_i}^{H_{n,1}(s_i)}\Big(\int_{x_i}^{x_{i+1}} c(s-G_m^{-1}(t))ds\Big) dt,$$
from which we see that 
\begin{eqnarray*} \textstyle
|\hat{T}_{n,m}-T_n|&\leq & 2 \textstyle \sqrt{n} \sum_{i=1}^{k-1}\Big| \int_{s_i}^{H_{n,1}(s_i)}\Big(\int_{x_i}^{x_{i+1}} |G_m^{-1}(t)-G^{-1}(t)|ds\Big) dt\Big|\\
&\leq &\textstyle 4M \|\alpha_{n,1}\| \sup_{0\leq t\leq 1} \|G^{-1}(H^{-1}_{m,2}(t))-G^{-1}(t)\|\to 0
\end{eqnarray*}
in probability, since $G^{-1}$ is continuous and $\sup_{t\in (0,1)} |H^{-1}_{m,2}(t)-t|=\sup_{x\in (0,1)} |H_{m,2}(x)-x|\to 0$ in probability.
Similar statements are true for $T_n^+$ and $T_n^-$. \FIN
}
\end{Nota}

\medskip
\noindent
\textbf{Proof of Theorem \ref{asymptotics}.}   
Convergence in the $L_2-$Wasserstein distance sense is characterized through weak convergence plus convergence of second order moments. Therefore the a.s. consistency $\varepsilon_{\mathcal W_2}(F_n,G_m)\convs \varepsilon_{\mathcal W_2}(F,G)$  essentially follows from the strong law of large numbers (see \cite{Cuesta92} for details and more general results).
For the asymptotic law, we write $\left(\varepsilon_{\mathcal W_2}(F_n,G_m)-\varepsilon_{\mathcal W_2}(F,G)\right)
=(\varepsilon_{\mathcal W_2}(F_n,G_m)-\varepsilon_{\mathcal W_2}(F,G_m))+
(\varepsilon_{\mathcal W_2}(F,G_m)-\varepsilon_{\mathcal W_2}(F,G))$. By Theorem \ref{casoL2} and Remark \ref{Notadosmuestras} arguing  as in the proof of Corollary \ref{indexL2}
we see that $\sqrt{n}(\varepsilon_{\mathcal W_2}(F_n,G_m)-\varepsilon_{\mathcal W_2}(F,G_m))\to_w N\Big(0,\frac {\mbox{Var}(v_-(U))} {\mathcal{W}_2^8(F,G)}\Big)$.
A minor modi\-fication of the proof of Corollary \ref{indexL2} yields that 
$\sqrt{m}(\varepsilon_{\mathcal W_2}(F,G_m)-\varepsilon_{\mathcal W_2}(F,G))\to_w N\Big(0,\frac {\mbox{Var}(v_+(U'))} {\mathcal{W}_2^8(F,G)}\Big)$
with $v_+$ as in (\ref{funcionesv}) and $U'$ a U(0,1) law,
and also that $\sqrt{n}(\varepsilon_{\mathcal W_2}(F_n,G_m)-\varepsilon_{\mathcal W_2}(F,G_m))$ and $\sqrt{m}(\varepsilon_{\mathcal W_2}(F,G_m)-\varepsilon_{\mathcal W_2}(F,G))$
are asymptotically independent. The result follows. \FIN

 \end{document}